\documentclass[journal=jacsat,manuscript=article]{achemso}

\usepackage{graphicx}
\usepackage{lipsum} % Required to insert dummy text
\usepackage[version=4]{mhchem}
\usepackage{siunitx}
\usepackage{physics}
\usepackage{setspace}
\DeclareSIUnit\Molar{M}
\usepackage{fancyref}
\usepackage[sort&compress]{natbib}
\usepackage{multirow}

\newcommand{\PR}[1]{\ensuremath{\left[#1\right]}}

\author{Ely G. F. de Miranda}
\affiliation{Instituto de F\'isica, Universidade de S\~ao Paulo,
Rua do Mat\~ao 1371, 05508-090, S\~ao Paulo, S\~ao Paulo, Brazil}
\email{ely.miranda@usp.br}

\author{Lucas M. Cornetta}
\affiliation{Instituto de F\'isica Gleb Wataghin, Universidade Estadual de Campinas, Rua S\'ergio Buarque de Holanda, 777 - Cidade Universit\'aria, Campinas, S\~ao Paulo, Brazil}
\email{lucascor@unicamp.br}

\author{M\'arcio T. do N. Varella}
\affiliation{Instituto de F\'isica, Universidade de S\~ao Paulo,
Rua do Mat\~ao 1371, 05508-090, S\~ao Paulo, S\~ao Paulo, Brazil}
\email{mvarella@if.usp.br}

\title[]
  {Low energy electron interactions with resveratrol and resorcinol: anion states and likely dissociation pathways}

\abbreviations{IR,NMR,UV}

\begin{document}

\begin{abstract}
We report a computational study of the anion states of the resveratrol (RV) and resorcinol (RS) molecules, 
also investigating dissociative electron attachment (DEA) pathways. RV has well known beneficial effects in human health,
and its antioxidant activity was previously associated with DEA reactions producing H$_2$. Our calculations
indicate a valence bound state ($\pi^*_1$) and four resonances ($\pi^*_2$ to $\pi^*_5$) for that system. 
While the computed thermodynamical thresholds are compatible with DEA reactions producing H$_2$ at 0~eV, the well known mechanism involving
vibrational Feshbach resonances built on a dipole bound state should not be present in RV. Our results suggest that the shallow $\pi^*_1$
valence bound state is expected to account for H$_2$ elimination, probably involving $\pi_1^*$/$\sigma_{\text{OH}}^*$ couplings along the vibration dynamics. 
The RS molecule is also an oxidant and a subunit of RV. Since two close-lying
hydroxyl groups are found in the RS moiety, the H$_2$-elimination reaction in RV should take place at the RS site. 
Our calculations point out a correspondence between the anion states of RV and RS, and even between the thresholds.  Nevertheless, the absence of bound anion states in RS, indicated by our calculations, is expected to suppress the H$_2$-formation channel at 0~eV. One is lead to conclude that the ethene and phenol subunits in RV stabilize the $\pi^*_1$ state, thus switching on the DEA mechanism producing H$_2$.
%To that extent, RS could be viewed as a less computationally expensive prototype for the production of H$_2$. 

%on the low energy anion states of Resveratrol (RV) and Resorcinol (RS), employing both scattering and bound states techniques. RV produces beneficial effects in human health by preventing a variety of illnesses. The mechanisms underlying this bioactivity could be related to its fragmentation in the mitochondrial intermembrane space triggered by free electrons. Our results point out a valence bound state, three shape resonances and a mixed-character resonance for RV. The zero-energy thresholds reported elsewhere and the presently calculated resonance spectrum are consistent with the mass spectroscopy data, in particular, with the production of H$_2$ (reaction that could account for the antioxidant activity) at nearly zero energy which could proceed from a vibrational Feshbach resonances arising from a valence bound state.  We also investigated the RS subunit of the RV, which could be a less computationally expensive prototype for the production of H$_2$. We obtained two shape resonances and a mixed-character resonance. There is a correspondence between the anion states of RV and RS, and even between the thresholds, but the lack of anion bound states for the smaller molecules should suppress the H$_2$-formation channel at 0~eV.
\end{abstract}

\maketitle
\thispagestyle{empty}

\begin{spacing}{2.0}
\section{Introduction}

Life in our planet relies on two fundamental process: respiration and photosynthesis. They provide the means for the cellular energy production which is necessary for the living organisms~\cite{metzler2003biochemistry}. 
Both processes are inseparably linked with the presence of electrons moving through biological matter~\cite{ahmad2018biochemistry}. This balance is fulfilled along electron transfer pathways, in particular the mitochondrial electron transport chains (ETCs)~\cite{pelster2016mitochondrial}. However, reactive oxygen species (ROS), which are also produced along the mitochondrial ETC, are efficient oxidative substances capable of damaging the mithocondrial membranes, attacking DNA and causing mutations~\cite{costa2011role}. Molecular oxygen reduction generates the anionic superoxide, ${\rm O}_2 + e^- \rightarrow {\rm O}_2^{-\bullet}$,  which can give rise to oxidative stress~\cite{murphy2008mitochondria} leading to pathologies such as cardiovascular diseases, Alzheimer's and Parkison's~\cite{koopman2012monogenic,andreyev2005mitochondrial}. Nonetheless, ROS can also be beneficial to the cell, e.g., by producing oxidative damage to pathologies and taking part in cellular signaling~\cite{murphy2008mitochondria}. Protective mechanisms that neutralize ROS are linked with the generation of native antioxidants or enzymatic activity. Oxireductive balance perturbations in mithocondrial intermembrane space are therefore believed to trigger significant cellular damage~\cite{pshenichnyuk2018interconnections}.

%{\color{red} [Este trecho está meio confuso, parece que o ROS fica sendo mau, depois bom, depois mau. É preciso organizar melhor as ideias, e explicar melhor o que é signaling.]
%ROS are efficient oxidative substances capable of damaging the mithocondrial membranes, attacking DNA and causing mutations~\cite{costa2011role}. Nonetheless, ROS can also be beneficial to the cell, e.g., by producing oxidative damage to pathologies and taking part in cellular signaling~\cite{murphy2008mitochondria}. Self-defense mechanisms that neutralize ROS are linked with the generation of native antioxidants or include enzymatic protection. Therefore, it is believed that oxireductive balance perturbations into mithocondrial inter-membrane can trigger significant cellular damage~\cite{pshenichnyuk2018interconnections}.
%}

Polyphenolic antioxidants, such as flavonoids and spinochromes, are good electron acceptors and can interact with electrons under reductive conditions in cells~\cite{modelli2013gas}. Dissociative electron attachment (DEA) to multiple OH-substituted aromatic compounds has been found to produce neutral H$_2$~\cite{modelli2013gas}, which is an antioxidant species used clinically~\cite{ohsawa2007hydrogen}. These facts suggest that the antioxidant activity of polyphenolic compounds could be related to the electron-induced production of molecular hydrogen. The xenobiotic antioxidant compounds could be reduced in the mitochondrial intermembrane space by charge leaking from the ETC. The reduction can be viewed as the formation of transient negative ions (TNIs), also called resonances, that can initiate DEA reactions producing H$_2$ among other species. These mechanisms have been proposed to underlie the antioxidant activity of the polyphenolic compound resveratrol (RV)~\cite{base} and also the antipsoriatic activity of anthralin~\cite{pshenichnyuk2014dissociative}.

\begin{figure}[ht!]
\centering
\includegraphics[width=0.62\linewidth]{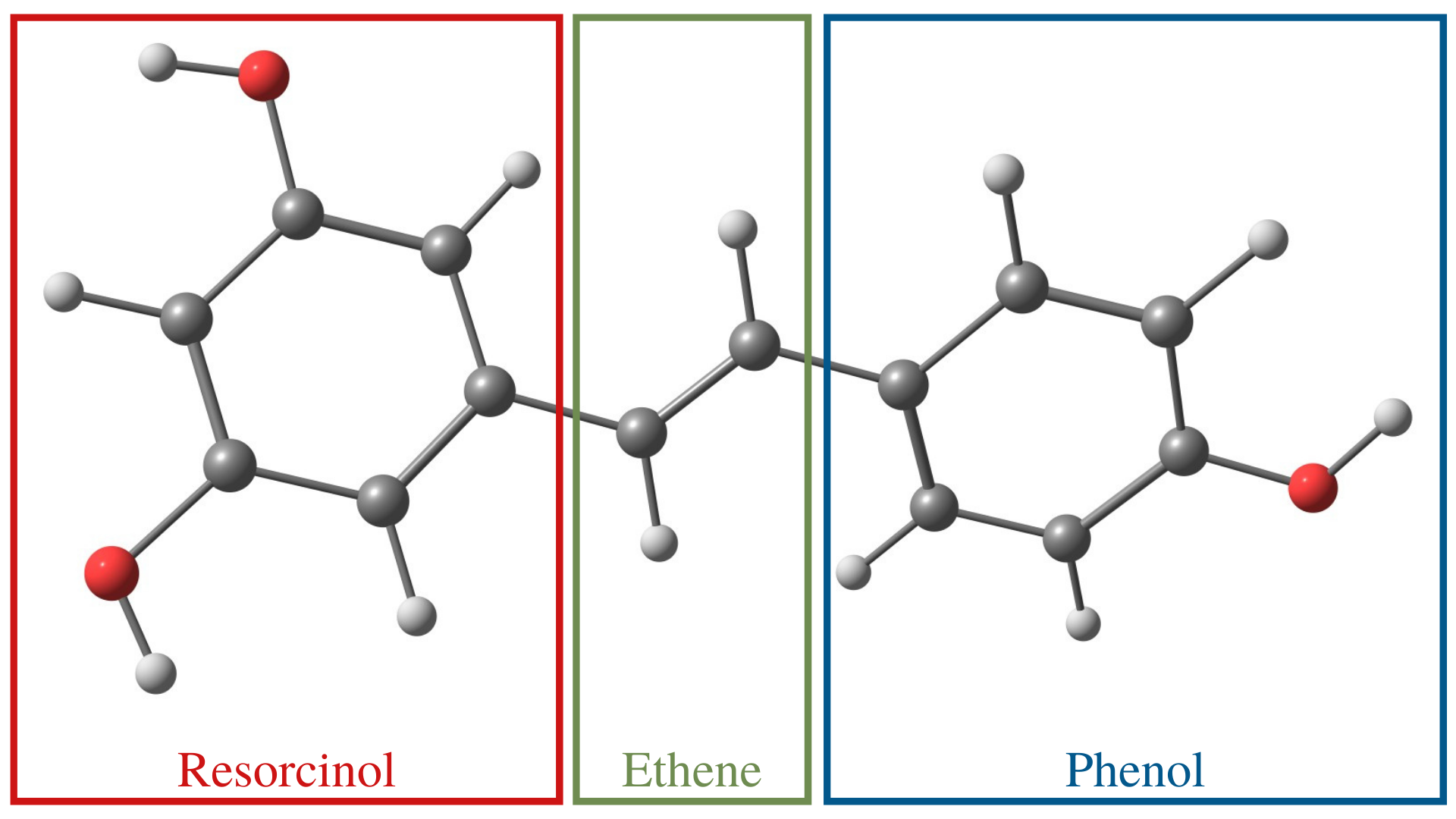}
\caption{Structure of the resveratrol molecule, with the oxygen atoms indicated in red, carbon in gray, and hydrogen in white. The resorcinol, ethene and phenol subunits are highlighted.}
\label{fig:frag}
\end{figure}

The RV molecule (3,5,4$^{'}$-trihydroxystilbene, shown in Fig.~\ref{fig:frag}) is a naturally occurring plant phytoalexin and a constituent of red wine. It is known for its strong antioxidant activity~\cite{baur2006therapeutic} and produces beneficial effects in human health by preventing a variety of illnesses, such as cancer, cardiovascular malfunction, neurodegenerative diseases, inflammation, and atherosclerosis~\cite{baur2006therapeutic}. The mechanisms underlying those
beneficial effects are not yet clear. Since H and H$_2$ can be produced by DEA to RV, we presently investigate the TNIs employing electron scattering calculations, and we also infer the possible dissociation pathways. A major difficulty to model the transient anion states of RV is the computational effort arising from the system size and lack of symmetry elements. The RV molecule has 29 atoms and can be considered large for scattering calculations. However, RV can be decomposed into resorcinol (RS), ethene and phenol (Ph) subunits, as indicated in Fig.~\ref{fig:frag}. The smaller subunits have been the subject of previous studies~\cite{de2012shape,szymanska2014dissociative}, so we presently consider RS, which is also an antioxidant \cite{arts2003critical,ortega2016substituent}. In view of the two  hydroxyl groups lying at meta positions with respect to each other in RS, 
H$_2$-elimination reactions in RV are likely to take place at the RS moiety.
To this extent, the smaller RS molecule could be a more suitable prototype for the DEA reaction producing molecular hydrogen, justifying the study of both the RS and RV molecules. We therefore investigate the bound and transient anion states of those molecules,
as well as the thermodynamical thresholds, in order to infer the DEA mechanisms for H- and H$_2$-elimination reactions.

This paper is organized as follows. In Sec. II we present the theoretical and computational methods for scattering calculations (Schwinger Multichannel Method with Pseudopotentials) employed for the anions characterizations. In Sec. III and IV we present the calculations and discuss our findings. Finally, the conclusions are summarized in Sec. V.

%The H elimination from hydroxyl groups would be expected in view of the strong polar character of the bonds, thus H-atom abstraction shall preferably occur in the PH and RS polyphenolic subunits (Fig.~\ref{fig:frag}). PH anionic states are already studied in literature~\cite{fenol}, in contrast to RS which should be the responsible for the H$_2$ formation of RV, since it has two hydroxyl groups. 

%Therefore, we employ scattering techniques to investigate the RV and RS anion states, in order to analyze the H$_2$ production through a DEA process. The subunit  investigation can present a good way to avoid a high computational effort, allowing an establishment of a methodology for the study of complex molecules. The present investigation may shed some light on the mechanism of biochemical reactions under reductive conditions and the therapeutic effects of polyphenolic compounds.

\section{Computational Procedures}
\label{two2}

\begin{figure}[ht!]
\centering
\includegraphics[width=0.82\linewidth]{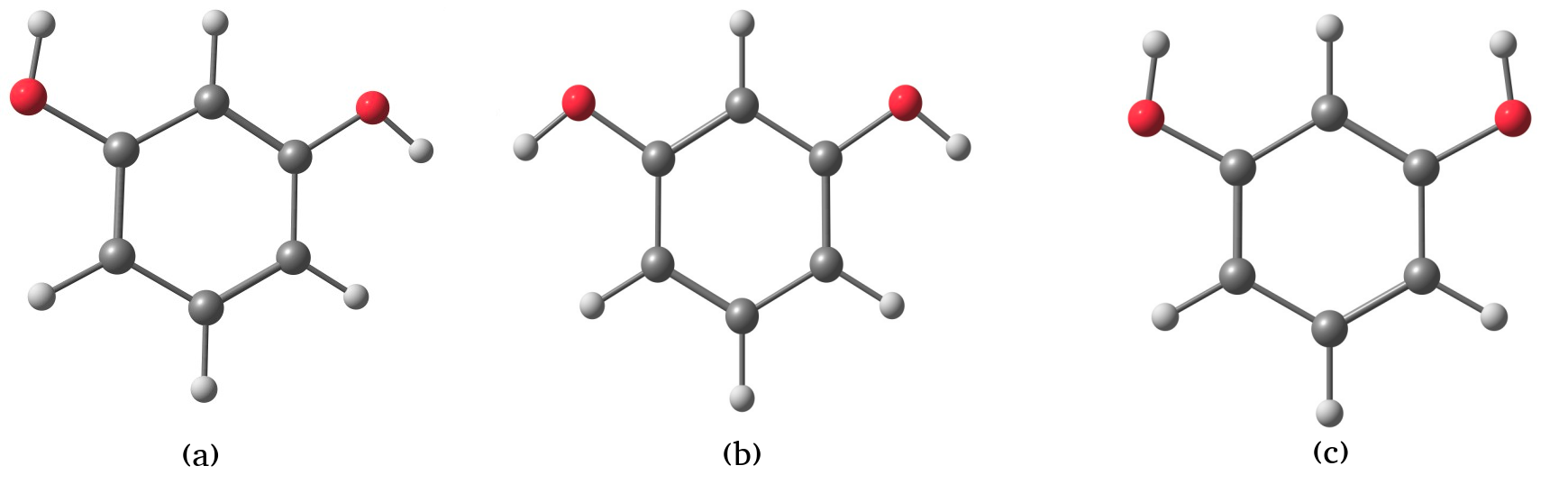}
\caption{Planar conformers of the resorcinol molecule, labeled (a) to (c). The oxygen atoms are indcated in red, carbon in gray and hydrogen in white.}
\label{fig:conf}
\end{figure}

All scattering calculations were performed in the fixed-nuclei approximation. Equilibrium geometries and vibrational analysis for the neutral molecules were performed with density functional theory (DFT), employing the $\omega$B97XD functional and the 6-311++G(d,p) basis set, as implemented in the Gaussian09 package~\cite{g09}.
Three planar conformers were identified for the RS molecule, differing by the orientation of the hydroxyl groups, as shown in Fig.~\ref{fig:conf}. The conformers (b) and (c) belong to the $C_{2v}$ symmetry point group, while the conformer (a) belongs to the $C_S$  group. 
Based on the free energies at room temperature ($k_BT\approx26$ meV), we assign the conformer (b) as the most stable one, lying 2 meV and 28 meV below the (a) and (c) forms, respectively. 
Therefore, an admixture of the three conformers should be present in gas-phase experiments justifying scattering calculations for all of them. In contrast to the small differences in the free energy values, the dipole moment magnitudes are significantly affected by the orientation of the hydroxyl groups. We obtained $\mu=1.4$~D for the optimal structure (a), against $\mu=2.5$~D and $\mu=2.3$~D for conformers (b) and (c), respectively.

\begin{figure}[ht!]
\centering
\includegraphics[width=0.78\linewidth]{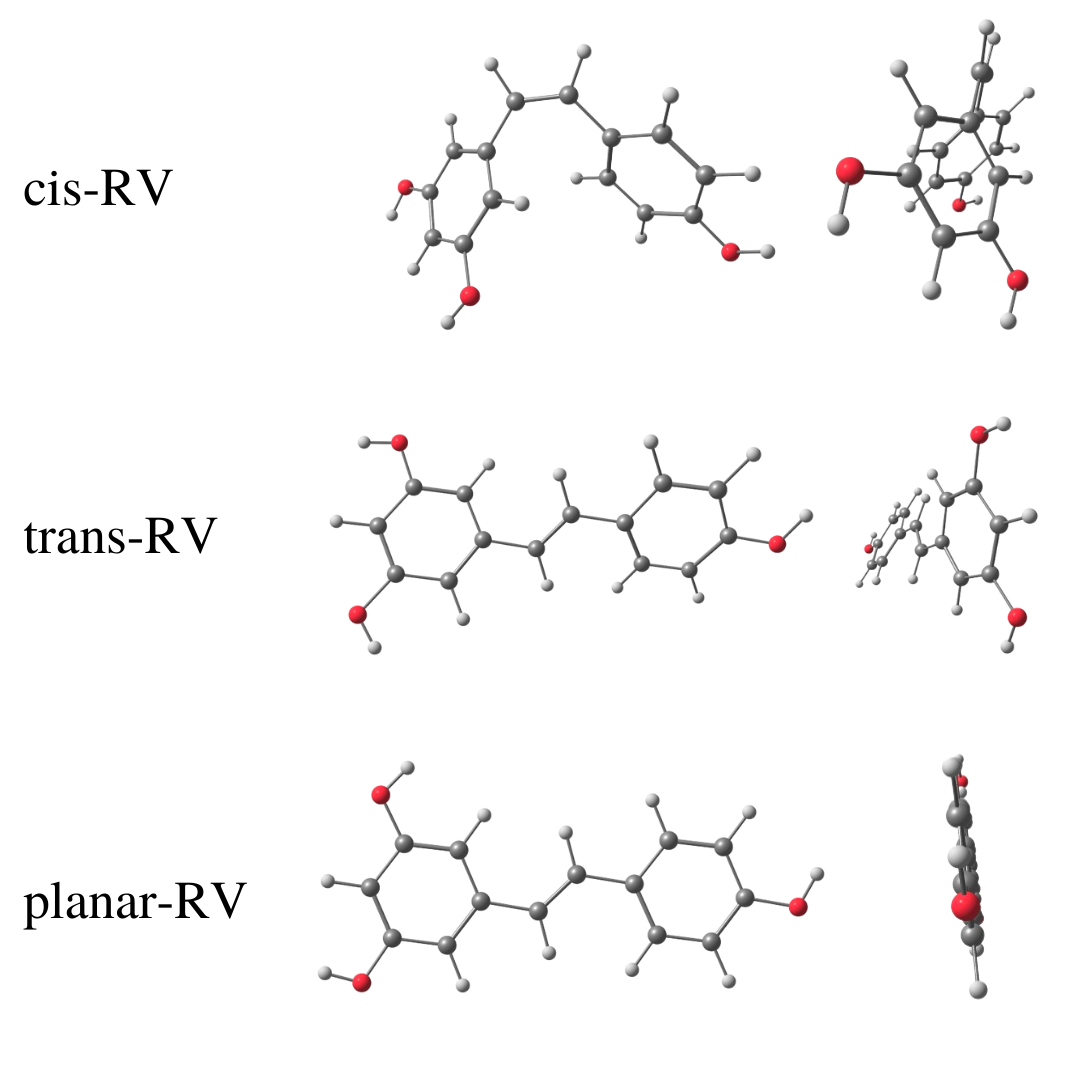}
\caption{Structures of the RV molecule computed with the $\omega$B97XD/6-311++G(d,p) method (front and side views). The oxygen atoms are represented in red color, carbon in gray, and hydrogen in white.}
\label{fig:rv}
\end{figure}

For the RV molecule, the existence of cis (c-RV) and trans (t-RV) isomers deserves a closer look. The relative ground state energies of the RV isomers, also calculated with the $\omega$B97XD/6-311++G(d,p) method, indicate that the t-RV isomer is significantly more stable than the cis counterpart, by 0.3 eV.
While the geometries of the cis and trans isomers are shown in Fig.~\ref{fig:rv}, we no further consider the cis form in view of the negligible Boltzmann populations. The geometry of the t-RV isomer was somewhat sensitive to the choice of the exchange-correlation functional. The $\omega$B97XD/6-311++G(d,p) calculations predict a
torsion angle between the aromatic rings around 15 degrees (see Tab.~S1). However, exploratory optimizations performed with the B3LYP/6-311++G(d,p) method indicate a smaller angle, around 4 degrees, in agreement with previously reported DFT/B3LYP results~\cite{base}.
%were also optimized at the $\omega$B97XD/6-311++G(d,p) level, 
%the results were more sensitive to the choice of the exchange-correlation functional. 
%The equilibrium geometry of t-RV indicates a torsion angle between the aromatic rings of about $\approx 15$ degrees (Tab.~S1 in the Supplementary Information).
%However, exploratory optimizations performed with the B3LYP functional indicate smaller a torsion of only $\approx 4$ degrees, which in turn agrees with previously reported DFT/B3LYP results~\cite{pshenichnyuk2015dissociative}. 
Since dispersion interactions are accounted for in the $\omega$B97XD functional, although not in B3LYP, the discrepancy in the torsion angle suggests that the interaction between the unsaturated rings would not be accurately described without dispersion corrections~\cite{tsuzuki2002origin,hwang2015important}. For the RS conformers, not having inter-ring $\pi$-$\pi$ interactions, the B3LYP and $\omega$B97XD functionals predict similar structures.

Despite the deviation from planarity arising from the torsion angle, the scattering calculations employed a planarized geometry of the t-RV isomer, referred to as p-RV (also computed with the $\omega$B97XD/6-311++G(d,p) method, see Fig.~\ref{fig:rv}). Previous scattering studies of nucleobases \cite{doi:10.1063/1.2424456,doi:10.1063/1.3675448,Nunes} used a similar procedure, since the planar structures allow for symmetry decomposition which reduces the computational effort. In addition, the signatures of the shape resonances become more evident in the symmetry-decomposed calculated cross sections. We estimated errors around 0.1eV between t- and p-RV resonance positions employing empirically corrected virtual orbital energies, according to Scheer and Burrow~\cite{escala}.

The scattering cross sections were computed with the parallel version of the Schwinger multichannel method implemented with the Bachelet-Hamann-Schlüter~(BHS) pseudo potentials (SMCPP)~\cite{santos2012low,bettega1993transferability}. 
The method was recently reviewed~\cite{da2015recent}, so we briefly outline a few relevant aspects. 
The scattering electronic wave function is given by
\begin{equation}
    \ket{\Psi^{(\pm)}_{\textbf{k}_i}}=\sum_\mu c_\mu^{(\pm)}({\textbf{k}_i}) \ket{\chi_\mu},
\end{equation}
where $\textbf{k}_i$ denotes the wave vector of the incident electron and $+(-)$ stands for the boundary condition associated with outgoing (incoming) spherical waves~\cite{joachain1975quantum}. 
The method uses square-integrable functions to build the trial set $\{\chi_\mu\}$, which is very convenient from the computational point of view.
The basis vectors $|\chi_\mu\rangle$ are referred to as configuration state functions (CSFs), and correspond to spin-adapted (${N + 1}$)-particle Slater determinants built on the closed-shell ground state of the neutral target, comprising $N$ electrons. The latter reference state is described at the restricted Hartree-Fock (RHF) level, employing sets of Cartesian Gaussian basis functions. The 5$s$5$p$2$d$ basis sets described by Bettega {\it et al}\cite{bettega1993transferability,bachelet1982pseudopotentials} were used for the carbon and oxygen atoms, while the 4$s$ basis set reported by Dunning~\cite{dunning1970gaussian} was used for the hydrogens, amounting to a total of 274 basis functions for each RS structure and 580 for the p-RV structure. The RHF calculations were performed with the GAMESS package~\cite{ishimura1993gamess}.

The scattering calculations were restricted to the elastic channel, which is a reasonable approximation for electron energies below the excitation threshold. Two models were explored to construct the CSF space.
In the static-exchange (SE) approximation, the target is kept frozen in the ground state $\ket{\Phi_0}$, while the unoccupied (virtual) orbitals are used as scattering orbitals $\ket{\phi_m}$, i.e., $|\chi_m\rangle = \mathcal{A}_{N+1} \ket{\Phi_0} \otimes \ket{\phi_m},$ where $\mathcal{A}_{N+1}$ is the anti-symmetrization operator. This approximation does not recover the response of the target electrons to the incoming electron.
%, being a reasonable approximation for collision energies above $\sim20$ eV.
Even though the SE approach retrieves the physics of shape resonances, at lower energies the dynamical response of the target electrons, referred to as correlation-polarization effects, should be accounted for. 
In the static-exchange plus polarization (SEP) approximation, virtual excitations of the target are allowed, so the CSF space is augmented with configurations of the type $|\chi_{im}\rangle = \mathcal{A}_{N+1} \ket{\Phi_i} \otimes \ket{\phi_m}$, in which $\ket{\Phi_i}$ represents a singly excitation of the target and $\ket{\phi_m}$ is a scattering orbital. 
In this approximation, long-range polarization (induced dipole moment) and short-range correlation are taken into account. 
The SEP functions require the choice of three orbitals, two of which are associated with the excitation of the target (hole and particle) and the third being the scattering orbital. The CSF space is build according to the energy criterion proposed by Kossoski and Bettega~\cite{kossoski2013low}. All single-particle orbitals whose energy eigenvalues satisfy $\epsilon_{\text{scat}}+\epsilon_{\text{part}}-\epsilon_{\text{hole}}<\Delta$, are taken into account, where $\epsilon_{\text{scat}}$, $\epsilon_{\text{part}}$, and $\epsilon_{\text{hole}}$ are the energies of the scattering, particle, and hole orbitals, respectively, while $\Delta$ is a cutoff. 
Finally, modified virtual orbitals (MVOs)~\cite{bauschlicher1980construction} were used as particle and scattering orbitals. They were generated from cationic Fock operators with charge $+6$ for RS and $+8$ for RV.

The integral cross sections (ICSs) of a-RS and p-RV were decomposed into the A$^{'}$ and A$^{''}$ components of the C$_S$ group, while for the b-RS and c-RS the ICSs are decomposed into the four components of the C$_{2v}$ group, namely A$_1$, A$_2$, B$_1$ and B$_2$.
In the SEP approximation for a-RS, we employed $\Delta =-0.95$ Hartree for the A$^{'}$ and A$^{''}$ components, generating 15405 and 15228 configurations.
Similarly, for b-RS (c-RS), we employed $\Delta =-0.85$ Hartree for 
the A$_1$ component, resulting in 9978 (10053) configurations, and $\Delta =-1.00$ Hartree for A$_2$, B$_1$ and B$_2$ components, resulting in 6619 (6633), 6682 (6724) and 6654 (6654) configurations, respectively.
For the case of p-RV, only the A$^{''}$ component  was considered, since we do not expect resonance signatures in the A$^{'}$ ICS component. In this case, we employed $\Delta =-0.95$ Hartree resulting in 17969 CFSs. 

While only elastic collisions were taken into account, some insight into the inelastic channels can be gained from bound state calculations. The lowest lying excited states of the neutral RS molecule were explored with the CASPT2 method ~\cite{roos1996applications}. These calculations were performed with the
OpenMOLCAS software~\cite{fdez2019openmolcas}, and the active space is described in the supplementary information (SI). Finally, DEA reaction thresholds were estimated with the composite G4(MP2) method~\cite{curtiss2007gaussian}, as implemented in the Gaussian09 package.
The thresholds were estimated as the difference between the G4(MP2) energies of products, including the anionic fragment, with respect to the energy of the neutral reactant (see Sec.~IV).

\section{Results}

\begin{figure}[ht]
\centering
\includegraphics[width=1\linewidth]{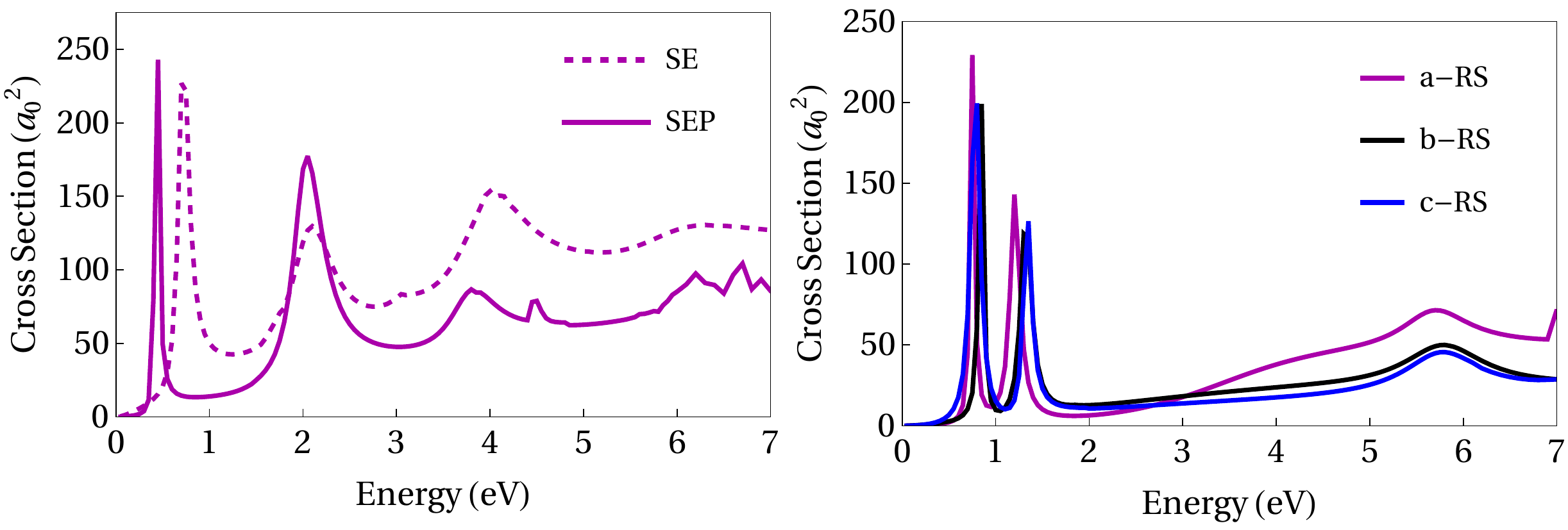}
\caption{Left panel: A$^{''}$ integral cross section of neutral p-RV calculated in SE - dashed line - and SEP - filled line - approximations; Right panel: A$_2$ plus B$_1$ (for b and c-RS) and A$^{''}$ (for a-RS) integral cross section for elastic electron scattering by the a (magenta line), b (black line) and c (blue line) structures of RS.}
\label{fig:scatt_comp}
\end{figure}

As expected, 
only $\pi^{*}$ resonances have clear signatures in the calculated ICSs, for both RV and RS. The corresponding peaks appear in the A$_2$ and B$_1$ components of $C_{2v}$ molecules (b- and c-RS), while in the A$^{''}$ component of $C_{S}$ molecules (a-RS and p-RV). 
It is worth noting that $C_S$ is a subgroup of $C_{2v}$, so the sum of the A$_2$ and B$_1$ components correspond to A$^{''}$. 
In the upper panel of Fig.~\ref{fig:scatt_comp}, we show the SEP-level ICSs obtained for a-RS (A$^{''}$ component), as well as b-RS and c-RS (A$_2$+B$_1$ components).
We observed three $\pi^*$ shape resonances for all geometries.
These are labeled $\pi^*_1$ to $\pi^*_3$, in order of increasing energy, according to the virtual orbitals that accommodate the additional electron. Respectively for a-, b- and c-RS, the anion states are located at 0.75, 0.84 and 0.78~eV ($\pi_1^{*}$); 1.21, 1.33 and 1.34~eV ($\pi_2^{*}$); and 5.70, 5.79 and 5.78 eV ($\pi_3^{*}$). The
positions obtained for the three structures are in good agreement among them. 
The excitation threshold of the first excited state of the target molecule was estimated as $\approx3.9$ eV for all isomers, 
indicating that the $\pi^{*}_3$ resonance might have a mixed shape and core-excited character.

\begin{figure}[ht]
\centering
\includegraphics[width=0.67\linewidth]{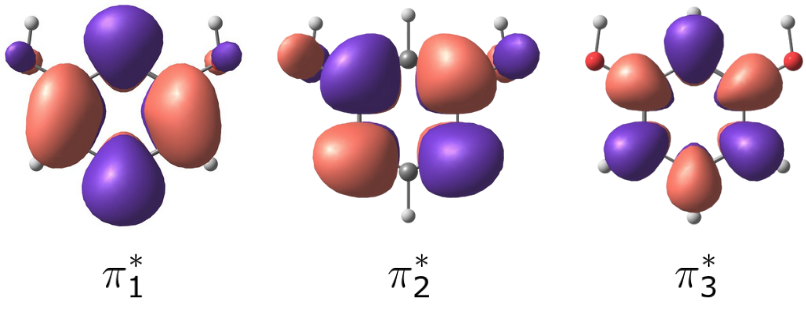}
\caption{$\pi^*$ type orbitals of RS obtained from the pseudo-eigenstates of the scattering Hamiltonian. The orbitals are linear combinations of the MVOs employed in the scattering calculations.}
\label{fig:hcomp}
\end{figure}

The SMCPP method allows for the diagonalization of the scattering Hamiltonian represented in the square-integrable CSF basis. 
This procedure generates a set of pseudo-states, since the actual spectrum is not discrete, but it provides insight into the resonance orbitals. The orbital plots shown in Fig.~\ref{fig:hcomp} were obtained from the SEP-approximation pseudo-eigenstates, although projecting them on the SE subspace. The properly
renormalized projections of the eigenstates can be recast as linear combinations of the MVOs, which represent the resonant orbital (occupied by the
attached projectile).
%by the projection of the SEP scattering wave function into the SE basis spawned subspace, i.e the scattering orbitals were analyzed as linear combinations of the MVOs. 
Even though we only show orbitals for the c-RS structure, similar results were obtained for the other RS isomers. The calculated positions and widths of the  resonance states are summarized in Tab.~\ref{tab:comp}. The SEP-approximation results for the RS structures are generally in good agreement, except for the width of the $\pi^*_1$ anion state. The 50-meV difference can nevertheless be viewed as small in absolute value.
The widths were obtained from standard local-approximation fits of Breit-Wigner profiles to the calculated eigenphase sums, so the origin of the discrepancy is unclear. 

%\color{red}We expect a discrepancy is for the $\pi^{*}_3$ state once it is typical of $\pi^*$ resonances having mixed shape and core-excited character and is related to the neglect of electronic excitation channels in the scattering calculations~\cite{barbosa2013shape,palihawadana2011low}. Despite the fact that the excited states are described by single excitations in our calculations, the threshold for the first triplet estimated with the CASPT2 method (3.9 eV) indeed suggest the possibility of a mixed-character resonance. The oscillations in the cross sections are spurious resonances that also arise from the elastic scattering approximation in the SEP calculations, i.e., for treating energetically open channels as closed.
%\color{black}

\begin{table}[ht]
\setlength{\tabcolsep}{10pt}
\renewcommand{\arraystretch}{1.4}
\centering
\caption{Resonance positions and widths (given in parenthesis), in units of eV, for a-, b-, and c-RS, as well as p-RV. SEP results are reported for RS, while SE and SEP for RV. The negative energy value indicates a bound state of the anion.}
\begin{tabular}{lccccc}
\hline \hline
System & $\pi_1^*$   & $\pi_2^*$   & $\pi_3^*$ & $\pi_4^*$   & $\pi_5^*$   \\ \hline
a-RS & 0.75 (0.05) & 1.21 (0.12) & 5.70 (0.91) & - & - \\
b-RS & 0.84 (0.06) & 1.33 (0.13) & 5.79 (1.02) & - & -\\
c-RS & 0.78 (0.11) & 1.34 (0.11) & 5.78 (1.02) & - & -\\ \hline 
p-RV (SE) & 0.72 (0.13) & 2.09 (0.54) & 4.02 (0.91) & 6.10 (2.12) & 10.24 (2.46) \\
p-RV (SEP) &  -0.07 & 0.44 (0.05) & 2.05 (0.39) & 2.05 (0.39) & 3.81 (0.56) \\ \hline \hline
\end{tabular}
\label{tab:comp}
\end{table}

%\begin{figure}[h]
%\centering
%\includegraphics[width=0.9\linewidth]{images/scatt_resv.pdf}
%\caption{$A''$ integral cross section of neutral p-RV calculated in SE - dashed line - and SEP - filled line - approximations.}
%\label{fig:choque-resv}
%\end{figure}

%\begin{table}[h]
%\centering
%\caption{Resonance positions (in eV) and ionization widths (in eV), given in parenthesis, for the p-RV geometry calculated with SE and SEP approximations. The negative position indicates a bound anion state.}
%\begin{tabular}{ccc}
%\hline \hline
%Orbitals    & SE    & SEP     \\ \hline
%$\pi_1^{*}$ & 0.72 (0.13)  & -0.07   \\
%$\pi_2^{*}$ & 2.09 (0.54)  & 0.44 (0.05)     \\
%$\pi_3^{*}$ & 4.02 (0.91)  & 2.05 (0.39)    \\
%$\pi_4^{*}$ & 6.10 (2.12)  & 2.05 (0.39)      \\
%$\pi_5^{*}$ & 10.24 (2.46) & 3.81 (0.56)      \\
%\hline \hline
%\end{tabular}
%\label{picosresv}
%\end{table}

\begin{figure}[ht]
\centering
\includegraphics[width=1\linewidth]{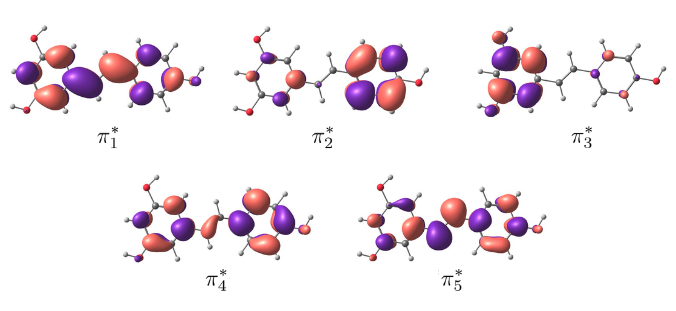}
\caption{p-RV $\pi^*$ type orbitals obtained from the pseudo-eigenstates of the scattering Hamiltonian. The orbitals are linear combinations of the MVOs employed in the scattering calculations.}
\label{fig:pseudo}
\end{figure}

The lower panel of Fig.~\ref{fig:scatt_comp} shows the A$^{''}$ component of the ICS for p-RV. The SE calculations point out five $\pi^*$ shape resonances around 0.72~eV ($\pi_1^*$), 2.09~eV ($\pi_2^*$), 4.02~eV ($\pi_3^*$), 6.10~eV ($\pi_4^*$) and 10.24~eV ($\pi_5^*$) (see also Tab.~\ref{tab:comp}). 
The interpretation of the SEP results for this system turned out to be more difficult, since we only identified three $\pi^*$ shape resonances, at 0.44~eV, 2.05~eV and 3.81~eV. The SEP spectrum could be clarified by inspecting the pseudo-eigenstates of the scattering Hamiltonian, as described above . The lowest-lying $\pi^*_1$ state becomes bound with respect to the neutral molecule by 70~meV, in consistency with previous results~\cite{base}. 
The 0.44-eV peak can be assigned to the $\pi^*_2$ resonance, while the $\pi^*_3$ and $\pi^*_4$ states merge into a single peak around 2.05 eV. The overlap between these resonances explain the apparently absent peak in the SEP results.
Lastly, the signature of the $\pi^*_5$ state, expected to have mixed character, is found around 3.8~eV. 
In general, a balanced description of polarization effects among several resonances belonging to the same irreducible representation is a challenging task in SEP calculations. Since there are five anion states in the A$^{\prime\prime}$ symmetry component of RV, our SEP results might not be properly balanced. It should be clear that going beyond the SEP approximation for a system as large as RV would not be a trivial computational task. The $\pi^*$ orbital plots obtained from the SEP-approximation pseudo eigenstates projected on the SE space are shown in Fig.~\ref{fig:pseudo}.

%When more than a single resonance is found in the same symmetry component, it is not a simple task to obtain a balanced description of the correlation-polarization effects~\cite{kossoski2013low}, so RV is a very challenging system in this regard. 
%The $\pi_1^*$ resonance became an anionic bound state with the vertical energy of -0.07~eV, as expected~\cite{base}. 
%Its binding energy was obtained from the diagonalization of the scattering Hamiltonian in the CSF basis. 
%We assign the peak at 0.44 eV to be associated with the $\pi^*_2$ resonance.
%The $\pi^*_3$ and $\pi^*_4$ lie very close to each other, and their signatures merge into a single peak (2.05 eV), which explains the apparent absence of a fourth peak.
%Lastly, the signature of the $\pi^*_5$ state, expected to have mixed character, is found at 3.8~eV. The orbital plots shown in Fig.~\ref{fig:pseudo} were obtained from the SEP-approximation pseudo-eigenstates. 
%Our results seem to be reasonable and the accuracy of the method when compared to electron transmission data is typically around 0.3~eV, such that the cost-benefit of a more expensive calculation could be questioned.

%??? Achei melhor tirar
%We can further analyze the RV anion states by decomposing the molecule into the RS and PH subunits. The PH molecule has three well-known shape resonances~\cite{de2012shape,jordan1976electron,khatymov2003phenol}. The two resonances of RV localized on a single subunit clearly correspond to the resonances of the subunits, i.e., the $\pi^*_2$ ($0.44$~eV) and $\pi^*_3$ ($2.05$~eV) states of RV correlate with the $\pi^*_1$ state in PH ($0.91$~eV) and the $\pi^*_2$ ($1.21$~eV) state in RS, respectively. The resonances could be expected to stabilize in RV with respect to the monomers since there is some degree of delocalization, and mutual polarization of the subunits.  This was not the case for the $\pi^*_3$ resonance in RV, corresponding to $\pi^*_2$ in RS, although this point was not further investigated.

\section{Discussion}
\label{sc:Discussion}

\begin{figure}[ht]
\centering
\includegraphics[width=0.85\linewidth]{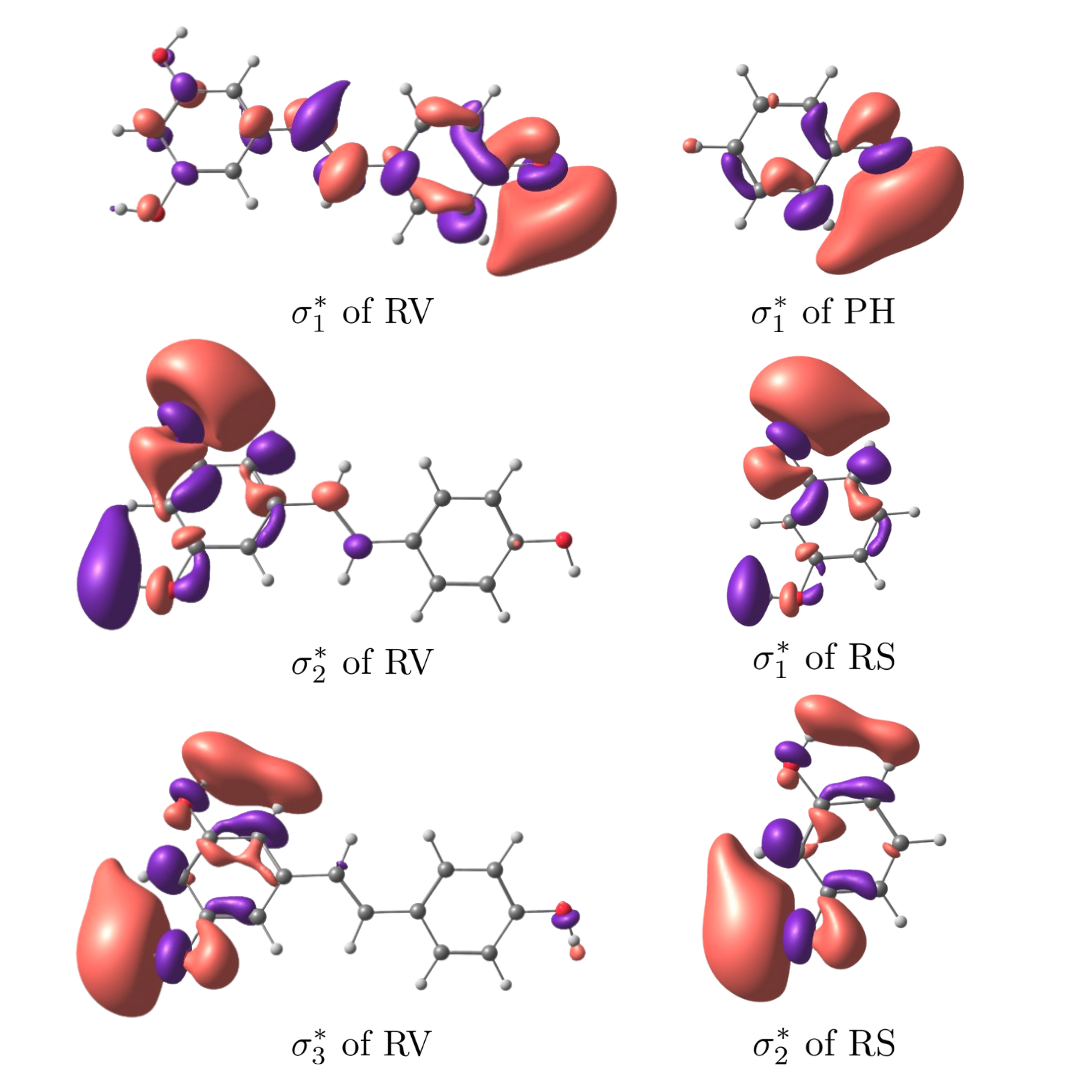}
\caption{$\sigma^*$ virtual orbitals obtained with the $\omega$B97XD/6-31G* for the RV, RS and Ph molecules.}
\label{fig:sigmas}
\end{figure}

DEA to RV was investigated experimentally~\cite{base,pshenichnyuk2018interconnections}. While the target molecules were 
evaporated at 170$^{\circ} $~C, and the collision cell kept at 180$^{\circ} $~C to avoid condensation, the c-RV isomer is not expected to
have a significant Boltzmann population, as discussed above. The main fragments observed in the energy- and mass-resolved measurements are
summarized in Tab.~\ref{tab:exper}, along with the peak energies and relative intensities.
%DEA to RV was investigated experimentally~\cite{pshenichnyuk2015dissociative}. It was found that the mass of anionic fragments are observed as a function of the incident energy of the electrons. It is worth mentioning that the RV molecule was evaporated at 170$^{\circ} $~C and the collision cell was kept at 180$^{\circ} $~C to avoid condensation. Even for this high temperature, a relevant population of the c-RV isomer is not expected. A more detailed description of the experimental procedure can be found at Ref.~\cite{pshenichnyuk2018interconnections}. The main fragments reported by in Ref.~\cite{pshenichnyuk2015dissociative} are reproduced in Tab.~\ref{tab:exper}, along with the energies of the signal peaks and their relative intensities. We employed the notation $[\text{M}-\text{X}]^-$ to denote the anion fragment formed by the release of the neutral fragment X from the parent M$^-$ molecular anion. For instance, hydrogen eliminations correspond to the reaction ${\rm M} + e^- \rightarrow [{\rm M}-{\rm H}]^- + {\rm H} \;.$ 
The calculated energy of the $\pi^*_2$ state, 0.44 eV, is compatible with the DEA peak for the elimination of two hydrogens (0.6 eV). The nearly degenerate resonances, $\pi^*_3$ and $\pi^*_4$, are expected to give rise to the most intense DEA signal (1.2 eV), corresponding to the H-elimination reaction. The 
disagreement in energy is another indication that our scattering calculations overestimate the resonances positions (2.0 eV). Finally, the DEA peaks around 4 eV are compatible with the $\pi^*_5$ resonance, which was predicted at 3.8 eV (SMCPP). In this case the SMCPP result seems in better agreement with the data, 
but it could be to some extent fortuitous, since core-excited anion states may account for the signal around 4.0 eV, at least partly. The abstraction of H atoms mediated by $\pi^*$ resonances should also involve $\sigma^{*}$ anion states with anti-bonding character on the 
polar OH groups. Those resonances 
typically do not have clear signatures in the scattering cross sections~\cite{santos2012low,cornetta2017transient,kossoski2014shape}, although indirect evidence
can be provided by virtual orbitals calculated with compact basis sets. As shown in Fig.~\ref{fig:sigmas}, we could find anti-bonding virtual orbitals with $\sigma^*_\text{OH}$ character and similar amplitudes in the Ph and RS subunits of the RV molecule.

\begin{table}[ht]
\setlength{\tabcolsep}{12pt}
\renewcommand{\arraystretch}{1.4}
\centering
\caption{Anion fragments observed in DEA experiments~\cite{base}. The fragments, peak energies (in eV) and relative intensities are shown 
(the asterisk indicates a low intensity).}
\label{tab:exper}
\begin{tabular}{lcc}
\hline \hline
Fragment                        & Energy & Intensity \\ \hline
M$^-$                           & 0.0    & 45        \\ \hline
{{[}M-H{]}$^-$}  & 1.2    & 100       \\
                                & 4.3    & 24        \\ \hline
{{[}M-2H{]}$^-$} & 0.0    & 17        \\
                                & 0.6    & *         \\ \hline
{{[}M-3H{]}$^-$} & 4.5    & 0.8       \\
                                & 9.5    & 1.3       \\ \hline
{[}M-(CH)$_2$OH{]}$^-$          & 8.7    & 0.8       \\ \hline \hline
\end{tabular}
\end{table}

%Unfortunately, we could not calculate dissociation thresholds for the RV molecule with the G4(MP2) method due to hardware limitations, although zero-energy thresholds obtained form DFT computations were reported in Ref.~\cite{pshenichnyuk2015dissociative}. In the following, we restrict the discussion to the lower-energy fragments since our calculations do not account for core-excited resonances.

\begin{figure}[ht]
\centering
\includegraphics[width=0.85\linewidth]{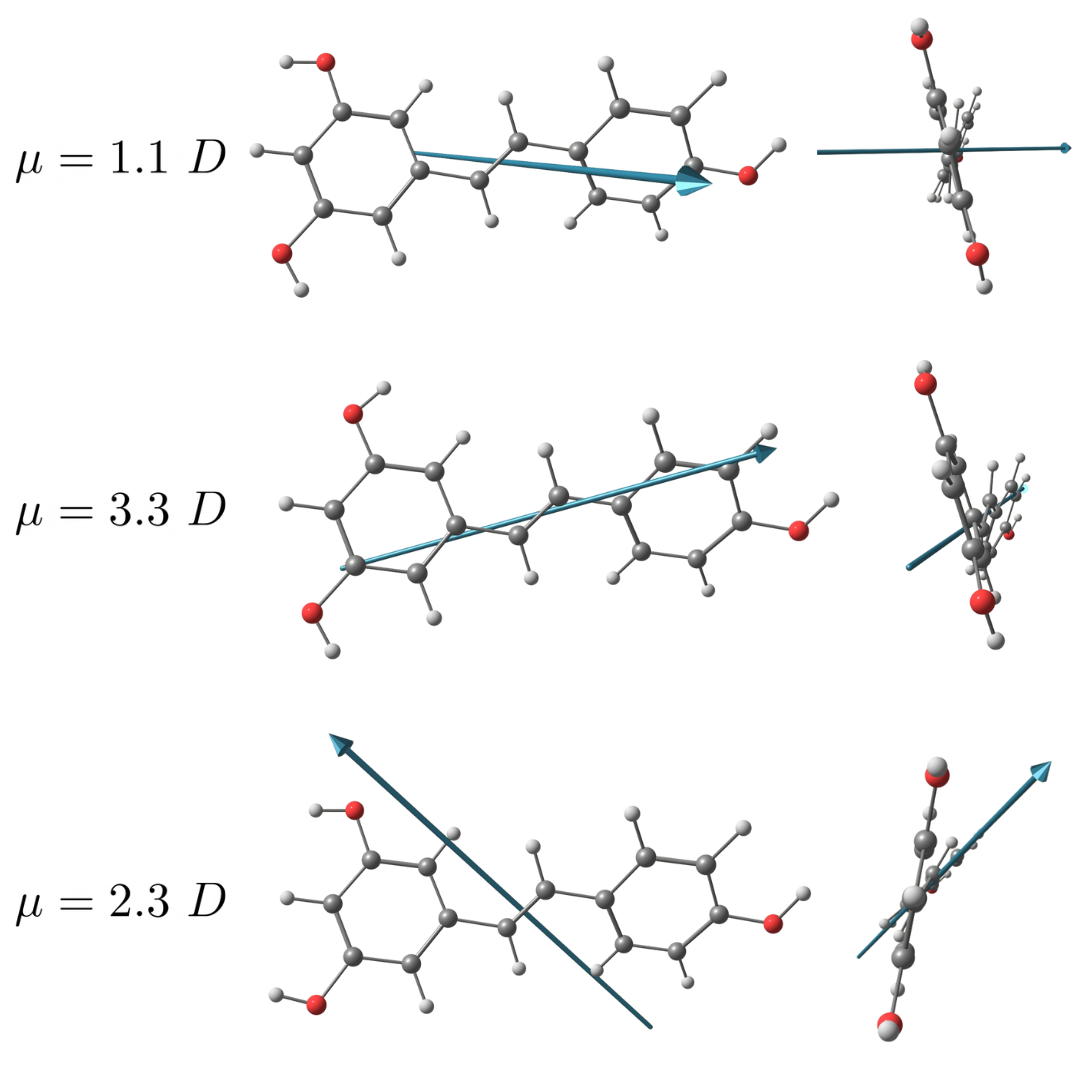}
\caption{Dipole moment vector and magnitude of the three RV isomers calculated with $\omega$B97XD/6-311++G(d,p).}
\label{fig:dip_RV}
\end{figure}

Turning attention to the DEA signals at 0~eV, one should remind that the dipole moment of the RV molecule can exceed $3.0$~D in case the hydroxyl groups in the RS subunit are properly aligned, as shown in Fig. \ref{fig:dip_RV}. We employed the aug-cc-pVDZ basis set augmented with a 6s6p set of diffuse functions, as suggested by Skurski \textit{et al.}~\cite{skurski2000choose}, to survey dipole-supported states in the RV conformers. We could only converge a DBS for the most
strongly polar form ($\mu=3.3$ D). The binding energy obtained at the MP2 level was very small, below 1 meV. Even if the DBS exists at all,
we believe the well known mechanism for H elimination~\cite{burrow2006vibrational}, initiated by vibrational Feshbach resonances (VFRs) built on a DBS, would not account for the 0~eV signal in RV. According to our $\omega$B97XD/6-311++G(d,p) calculations, the ground state of neutral RV has three OH stretch modes with energies close to 3920~cm$^-1$ (0.49~eV). In view of the nearly zero binding energy of the DBS, the $\nu_{\rm OH}=1$ VFRs would lie significantly above the observed DEA 
peak, around 0.5 eV.  In addition, the orientation of the dipole vector in the most strongly polar conformer favors H elimination from the Ph subunit, 
such that the formation of molecular hydrogen would be unlikely.
The direction of the dipole moment fovars H$_2$ elimination in a less polar conformation ($\mu=2.3$ D, see Fig. \ref{fig:dip_RV}), but we could not obtain
a DBS for this conformer. 

As mentioned above, the diagonalization of the scattering Hamiltonian indicates a bound $\pi^*_1$ state for the p-RV geometry. This result is confirmed by $\omega$B97XD/6-311++G(d,p) calculations for t-RV (binding energy of 50 meV), although not by the B3LYP/6-311++G(d,p) calculations. Since the DEA data at 0~eV is compatible with the existence of a bound anion state, we believe the shallow $\pi^*_1$ state should account for the low-energy DEA signals.
The binding energy ($\approx 50$~meV) is rather close to the thermal energy under experimental conditions ($k_BT=39$~meV), such that VFRs at 0~eV can arise from low-energy vibrational modes, ranging from $\sim 0$ to $\sim 50$ meV. 
Relaxation of the $\pi^*_1$ state is favored by the dense vibrational spectrum, in consistency with the
observation of the parent M$^-$ anion at 0~eV. Even though the pathway for H$_2$ elimination is not evident, 
it should involve $\pi^*_1$/$\sigma^*_{\rm OH}$ couplings along the vibration dynamics.

%In addition, since the dipole moment of the RV molecule can exceed $3.0$~D in case the hydroxyl groups in the RS subunit are properly aligned, a shallow DBS could also be expected. As a result, two bound anion states could give rise to vibrational Feshbach resonances which are consistent with the DEA data at 0 eV. The observation of the parent anion (M$^-$) suggests stabilization by vibrational energy redistribution, while the elimination of two hydrogen atoms should be favored by the formation of H$_2$, a very stable reaction product.

%The abstraction of one or more H atoms in DEA reactions should involve $\sigma^{*}$ resonances with anti-bonding character in the polar OH groups. This resonances typically do not have clear signatures in the scattering cross sections~\cite{cornetta2017transient,kossoski2014shape}. It is a common practice to look for indirect evidence of those $\sigma^*$ states in the virtual orbitals calculated with compact basis sets. As shown in Fig.~\ref{fig:sigmas}, not only we could find anti-bonding virtual orbitals with $\sigma^*_\text{OH}$ character, but they also have similar amplitudes as those calculated for the PH and RS subunits.

%As previously observed, the DEA reaction that produces H$_2$ has been claimed relevant for the antioxidant activity of RV~\cite{pshenichnyuk2015dissociative}. The RV molecule, especially in view of its size, is a very challenging system for scattering and even bound-state simulations, so we proposed RS as a prototype in the present study, since H$_2$ is likely to be produced in the RS subunit of RV, where two hydroxyl groups lie close to each other. 

\begin{figure}[ht]
\centering
\includegraphics[scale=0.37]{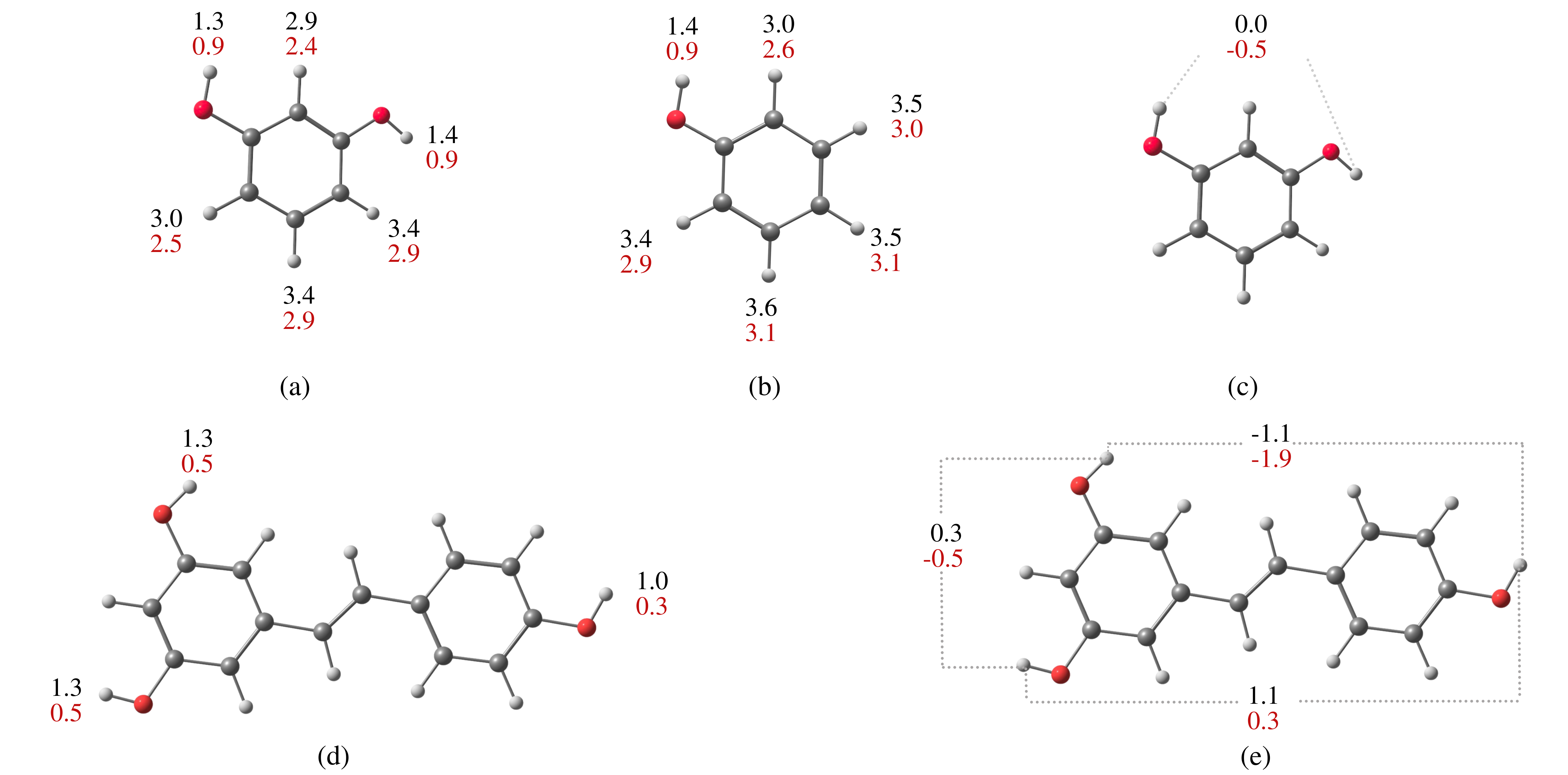}
\caption{Free energy dissociation thresholds (in units of eV) obtained for: (a) H abstraction from a-RS (a-RS+$e^-$ $\rightarrow$ $\PR{\text{(a)-RS  - H}}^-$+H$^\bullet$); (b) H abstraction from Ph ($\text{Ph}+e^-$ $\rightarrow$ $\PR{\text{Ph  - H}}^-+$H); (c) H$_2$ abstraction from a-RS (a-RS+$e^-$ $\rightarrow$ MBQ$^{\bullet -}$+H$_2$); (d) H abstraction from p-RV (p-RV+$e^-$ $\rightarrow$ $\PR{\text{p-RV  - H}}^-$+H$^\bullet$) and; (e) H$_2$ abstraction from p-RV (p-RV+$e^-$ $\rightarrow$ $\PR{\text{p-RV  - 2H}}^{\bullet -}$+H$_2$).  The black and red numbers indicate, respectively, the thresholds
at $T=0$ and $T=453.15$~K. The sites where the abstraction reactions take place are indicated by the position of the energy values in each panel.}
\label{fig:there}
\end{figure}

The calculated reaction thresholds for the H- and H$_2$-elimination channels are shown in Fig.~\ref{fig:there} for zero temperature (black) and $T=453.15$~K (red), 
which is consistent with the experimental condition. We show results for the a-RS conformer (results are similar for the other conformers),
Ph and RV. Our calculations corroborate that H$_2$ elimination at low energies should not take place at the Ph subunit, since the dissociation of the
CH bonds is energetically less favorable compared to the polar OH bonds. On the other hand, the thresholds for H abstraction from Ph and
a-RS are similar (though not discussed here, our results are consistent with with the DEA data for Ph~\cite{de2012shape,jordan1976electron,khatymov2003phenol}). The
present threshold estimates are generally in agreement with previous calculations~\cite{base}.

The formation of H$_2$ from RV, along with the meta-benzoquinone (MBQ) anion, 
has a nearly zero threshold at $0$~K, and the reaction becomes exothermic at higher temperatures. This is of course consistent with the 
formation H$_2$ at 0 eV. The calculated thresholds for H abstraction at 453~K are 0.3 eV (Ph subunit) and 0.5 eV (RS subunit). These estimates are also consistent with the observation of the [M-H]$^-$ fragment at 1.2~eV although not at 0~eV. Other fragments were observed at considerably higher energies
(see Tab. \ref{tab:exper}), so they cannot be interpreted based on our calculations.

%for a-RS conformer (other conformers shows equivalent results). The formation of H$_2$, along with the meta-benzoquinone (MBQ) anion, has a nearly zero threshold at $0$~K, and the reaction becomes exothermic at higher temperatures. While in this respect RS would be similar to RV, since the latter produces H$_2$ at 0 eV, the absence of anion bound states in RS should suppress DEA at 0 eV, since the reactions are expected to be initiated by vibrational Feshbach resonances. The dissociation thresholds for both RS structures are typically above the energy of the $\pi^*_1$ resonance (0.8 eV). Therefore, H elimination would not proceed from the formation of the $\pi^*_1$ state, although H$_2$ elimination is energetically allowed. Based only on the energetics, one could expect elimination of H and H$_2$ initiated by the $\pi^*_2$ resonance (1.4 eV). Of course DEA processes should also be initiated by higher-lying anion states, but we do not discuss those processes here.

As stated above, one of the goals of our work was to investigate if the smaller RS molecule would be a prototype for the H$_2$-elimination
reactions in RV. Molecular dynamics and reaction paths simulations would be considerably less demanding for the smaller system, apart
from the advantages discussed above for the scattering computations (symmetry decomposition, dimension of CSF space and polarization balance).
While the formation of H$_2$ would also be exotermic in RS, the reaction should be suppressed by the absence of bound anion states.
As stated above, the dipole moment magnitudes range from 1.4~D to 2.5~D, depending on the isomer, 
and our calculations do not indicate a DBS for RS (if the DBS exists at all, it should be very shallow). In addition, the $\pi^*_1$ state, which
is bound in RV, becomes a resonance in RS, lying around 0.7~eV. Since there are not bound anion states, the observation
of [M-2H]$^-$ fragments at 0~eV is not expected for RS. 
Our results are not in disagreement with the known antioxidant properties of RS, since mechanisms not based 
on DEA have been pointed out \cite{ortega2016substituent}. The differences between RV and RS actually suggest that the ethene and Ph subunits of
play a relevant role in the H$_2$-elimination reaction, although an indirect one. 
Those subunits switch on the H$_2$-elimination mechanism in RV, to the extent that they stabilize the 
$\pi^*_1$ state.

\section{Conclusions}

We reported a theoretical investigation of the low-energy anion states of the polyphenolic compound RV. This is a challenging system for electron scattering calculations in view of its size and low symmetry. Even though description of correlation-polarization effects is not well balanced in our SEP results, our study provides a consistent interpretation of the previously reported DEA measurements. In addition to four shape resonances, our results point out a shallow valence bound state $\pi^*_1$. This anion state, along with the computed dissociation thresholds, supports that H$_2$ elimination at 0 eV would be initiated by $\pi_1^*$/$\sigma_{\text{OH}}^*$ couplings. Since we could only converge a very shallow DBS for RV, we do not believe that H elimination at 0 eV is initiated by vibrational Feshbach resonances (VFRs) built on a DBS. The $\pi_1^*$ state is also responsible for the formation of the parent M$^-$ ion, also observed at 0 eV. The H-elimination reaction observed around 1.2~eV is triggered by two nearly degenerate shape resonances, $\pi^*_3$ and $\pi^*_4$.

%RV is a very challenging system for this kind of study in view of its size. While the polarization effects do not seem properly balanced between in the anion states in the SEP calculations, our results point out a valence bound state, three shape resonances and a mixed-character resonance. Resonances could trigger H abstraction in the RV molecule according to the dissociation threshold energy estimated with the G4(MP2) method. The zero-energy thresholds reported elsewhere and the presently calculated resonance spectrum are consistent with the DEA data. In particular, with the production of H$_2$ at nearly zero energy which could proceed from vibrational Feshbach resonances arising from the valence bound state. We suspect a DBS could also play a role, but we could not converge the calculation so far.

We also investigated the RS subunit, which could be a less computationally expensive prototype for the production of H$_2$, a reaction that could account for the antioxidant activity of RV, at least partly. We obtained two $\pi^*$ shape resonances and a mixed-character resonance from SEP-level scattering calculations. There is a correspondence between the anion states of RV and RS, and even between the dissociation  thresholds. However, the absence of bound anion states for the RS molecule is expected to suppress the H$_2$-formation channel at 0~eV. Our calculations therefore unveil the indirect part played by the ethene and Ph moieties of RV. They switch on the H$_2$-elimination mechanism initiated by $\pi_1^*$/$\sigma_{\text{OH}}^*$ couplings, since they stabilize the $\pi^*_1$ state.

%The inclusion of solvation effects in the anionic states is in perspective for future studies. We could expect the $\pi^*_1$ state to become more stable in water environment, such that RS could be a better prototype for H$_2$ production in this case. The DBS, on the other hand, should be less important in a polarizable dielectric environment such as water.

\section*{Acknowledgments}
E. G. F. M. acknowledges financial support from S\~{a}o Paulo Research Foundation (FAPESP), under grants No. 2021/09837-7, Brazilian National Council for Scientific and Technological Development (CNPq), under grant No. 131628/2019-4. 
L. M. C. acknowledges financial support from FAPESP, under grants No. 2020/04822-9 and No. 2021/06527-7.
M. T. do N. V. also acknowledges financial support (CNPq) (grant No. 304571/2018-0) and FAPESP (grant No.2020/16155-7).
The calculations were partly performed with HPC resources from STI, University of São Paulo.

%\subsection{SE result for RS}

%\begin{table}[h!]
%\centering
%\caption{Resonance energy (in eV) and ionization widths (in eV), given in parenthesis parentheses, for the (a)- and (c)-RS geometries calculated in SE and SEP approximations.}
%\begin{tabular}{ccccc}
%\hline \hline
% & \multicolumn{2}{c}{(c)-RS} & \multicolumn{2}{c}{(a)-RS} \\ \hline
% & SE               & SEP              &	SE	        & SEP         \\ \hline
%%$\pi_1^*$ & 3.15 (0.72)      & 0.78 (0.11)      & 	3.10 (0.65)	& 0.75 (0.05)         \\
%$\pi_2^*$ & 3.92 (0.59)      & 1.34 (0.11)      & 	3.88 (0.61)	& 1.21  (0.12)      \\
%$\pi_3^*$ & 9.95 (2.91)      & 5.78 (1.02)      &  9.89 (2.62)	&  5.70 (0.91)       \\ %\hline \hline
%\end{tabular}
%\label{tab:resoresp}
%\end{table}

\end{spacing}

\footnotesize{
\bibliography{bib}
}

\end{document}

\maketitle

\begin{spacing}{2.0}

%\section{$\omega$B97XD and B3LYP geometry optimization}
\section{Geometry optimization}

\begin{figure}[ht!]
 \centering
 \includegraphics[width=0.8\textwidth]{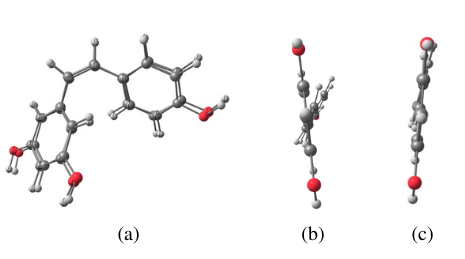}
 \caption{(a) Superimposed ground state geometries of the c-RV isomer optimized with the B3LYP and $\omega$B97XD functionals.
 The optimal geometries of the t-RV isomer obtained with the (b) $\omega$B97XD and (c) B3LYP functionals are also shown. All calculations employed
 the 6-311++G(d,p) basis set.
 %Overlap of the two c-RV optimized geometries performed with B3LYP and $\omega$B97XD; (b) t-RV optimized geometry performed with $\omega$B97XD; (c) t-RV optimized geometry performed with B3LYP.
 }
 \label{compcis}
\end{figure}

\begin{figure}[ht!]
 \centering
 \includegraphics[width=0.8\textwidth]{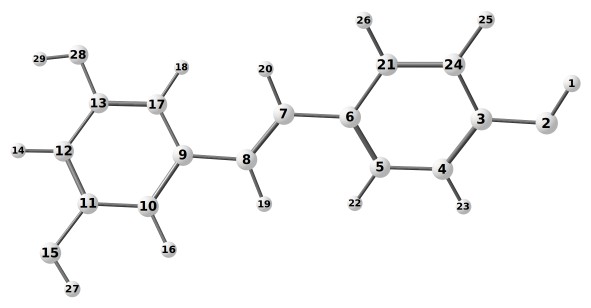}
 \caption{Atomic labels used in Tab.~\ref{diedro}}
 \label{figdie}
\end{figure}

The relative orientation of the two rings in t-RV significantly differs in the geometry optimizations performed with the B3LYP and $\omega$B97XD functionals, employing the 6-311++G(d,p) basis set in both cases. The optimal geometries are shown in Fig.~\ref{compcis} and Tab.~\ref{diedro}, where the dihedral angles indicate a nearly planar structure for the B3LYP functional. The labels used in the dihedral angles are defined in Fig.~\ref{figdie}. For the c-RV isomer, the geometries obtained with both functionals are similar (see Fig.~\ref{compcis}).

%As expected, the six-member rings significantly increases the contribution from long-range forces to short-range forces~\cite{tsuzuki2002origin,hwang2015important} with respect to the optimization calculation. For the planar-RV (p-RV) the geometries were practically equal).

\begin{table}[ht!]
\centering
\caption{Dihedral angles (in $^\circ$) for the t-RV isomer calculated with the $\omega$B97XD and B3LYP functionals. The atomic labels are given in Fig.~\ref{figdie}.}
\begin{tabular}{ccc}
\hline \hline
                & $\omega$B97XD         & B3LYP         \\ \hline
5-6-7-8           & 15.21          & 4.31          \\
7-8-9-10        & 159.89         & 173.86        \\
10-9-8-19     & 17.88          & 5.43          \\
5-6-7-20        & 166.59         & 176.28        \\ \hline \hline
\end{tabular}
\label{diedro}
\end{table}

\section{Excited states}

\begin{figure}[ht!]
\centering
\includegraphics[width=1\textwidth]{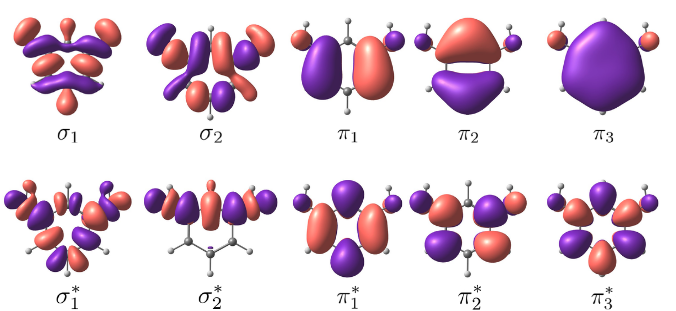}
\caption{Molecular orbitals included in the active space of the CASSCF calculations performed for the (c) isomer of the RS molecule. In the reference state, the five upper orbitals were considered doubly occupied and the five lower orbitals unoccupied.}
\label{fig:casscf1}
\end{figure}

\begin{figure}[ht!]
\centering
\includegraphics[width=1\textwidth]{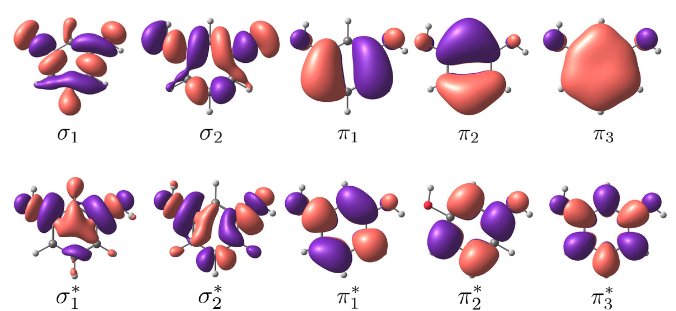}
\caption{Molecular orbitals included in the active space of the CASSCF calculations performed for the (a) isomer of the RS molecule. In the reference state, the five upper orbitals were considered doubly occupied and the five lower orbitals unoccupied.}
\label{fig:casscf2}
\end{figure}

The excitation spectra of the RS isomers were studied with the CASSCF/CASPT2 method and the ANO-L-VTZP basis set, employing the OpenMolcas software~\cite{fdez2019openmolcas}. These calculations can indicate the onset of electronic excitation in the scattering calculations and also possible parent states for core-excited resonances. The active space was built with 10 molecular orbitals. In the reference state, the 5 lowest-lying orbitals are doubly occupied and the remaining 5 orbitals are unoccupied. The molecular orbitals and their characters are indicated in Figs.~\ref{fig:casscf1} and~\ref{fig:casscf2} for the isomers (c) and (a), respectively (see the main text). Tab.~\ref{tab:caspt2} shows the results for the singlet and triplet states of both isomers. At least for the states considered here, the excitation energies are very close for the two conformers, with the lowest triplet states lying around 3.9 eV.

%\vspace{2cm}

\newpage

\begin{table}[ht!]
\centering
\caption{Vertical excitation energies (in eV) for the (c) and (a) isomers of RS obtained with the CASPT2 method. 
%Excitation energies of the anion molecules, expressed with respect to the neutral ground state, are also presented.
}
\label{tab:caspt2}
\begin{tabular}{cccc}
\hline \hline
\multicolumn{2}{c}{}                               & (c) & (a)  \\ \hline 
                           & $1^1 \pi\pi^*$                      & 4.43     & 4.41   \\
                           & $2^1 \pi\pi^*$ & 5.66     & 5.47    \\
\multirow{-3}{*}{Singlet} & $3^1 \pi\pi^*$ & 5.95     & 5.80       \\ \hline
                           & $1^3 \pi \pi^*$                     & 3.92     & 3.89      \\
                           & $2^3 \pi \pi^*$                     & 4.02     & 4.23      \\
\multirow{-3}{*}{Triplet} & $3^3 \pi \pi^*$                     & 4.49     & 4.42     \\ \hline
\hline
\end{tabular}
\end{table}

%\section{Configuration State Functions}

%The basis vectors $\{\ket{\chi_{\mu}}\}$ employed in static-exchange (SE) approximation are written as (), where .... Since the modified virtual orbitals (MVOs) correspond to the () part

%The SEP pseudo-eigenstates correspond to 

%by the projection of the SEP scattering wave function into the SE basis spawn subspace, i.e the scattering orbitals were analyzed as linear combinations of the MVOs.
%expanded as linear combinations of the MVOs.

\indent
\vspace{5cm}
%\footnotesize{
\bibliography{bib}
%}

\end{spacing}